\documentclass[cameraready]{Interspeech}
\usepackage{amsmath}
\usepackage{amssymb}
\usepackage{bm}
\usepackage{xcolor}
\usepackage{hyperref}

\title{ABSE-NET: A Lightweight Neural Model for Active Binaural Speech Enhancement in Open-Fit Hearing Aids}

\author[affiliation={1}, orcid={0000-0003-0226-917X}]{De}{Hu}
\author[affiliation={1}, orcid={0009-0005-9169-1066}]{Xue}{Du}
\author[affiliation={1}, orcid={0000-0003-4514-1619}]{Qingying}{Zhao}
\author[affiliation={2}, correspondingauthor, orcid={0000-0003-0955-4862}]{Qintuya}{Si}

\address{
    $^1$ College of Computer Science, Inner Mongolia University, China \\
    $^2$ College of Electronic and Information Engineering, Inner Mongolia University, China 
}

\email{cshood@imu.edu.cn, duxue@mail.imu.edu.cn, zhaoqingying@imu.edu.cn, siqty@imu.edu.cn}


\usepackage{comment}

\begin{document}

\maketitle

\begin{abstract}
    
    Open-fit hearing aids have attracted growing attention due to their superior wearing comfort. However, the open-fit design inevitably causes acoustic leakage into the ear canal, degrading the performance of existing binaural speech enhancement (BSE). To this end, we propose \textit{ABSE-NET}, an active BSE framework integrating active noise control (ANC) with BSE to jointly enhance target speech and suppress acoustic leakage. The ABSE-NET pipeline cascades a binaural MVDR (BMVDR) with a lightweight neural network (LNN). The former achieves a coarse BSE, whereas the latter simultaneously cancels acoustic leakage and compensates for BMVDR-induced distortion. The LNN uses an encoder-decoder with a feature fusion module, which includes frequency-time dependency learning and convolutional attention blocks. Unlike traditional BSE+ANC solutions via adaptive filtering, ABSE-NET needs no in-ear microphone in practical deployment. Experiments validate its superiority over state-of-the-art methods. Code repository: https://github.com/Bream101/ABSE-NET.
    
\end{abstract}

\section{Introduction}

\begin{figure}[t]
        \centering
        \includegraphics[width=1\columnwidth]{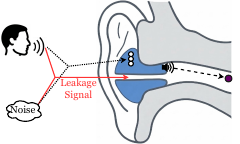}
        \caption{Acoustic leakage in semi-open-fit HAs. The external microphones (hollow circles) provide input for BSE, and the enhanced signal is played through the internal loudspeaker. However, external noisy signal leaks into the ear canal through the vent and corrupts the enhanced signal, which is detected by the error microphone (solid circle) placed deep inside the ear canal.}
        \vspace{-5pt}
        \label{fig:leakage path}
    \end{figure}

    Binaural speech enhancement (BSE) is a key component of modern hearing aids (HAs) and aims to deliver high-quality audio to hearing-impaired users \cite{dillon2012hearing,hearingloss2,hearingloss4}. 
    Currently, most BSE approaches are designed for closed-fit HAs, which cause the occlusion effect and may lead to discomfort during long-term use \cite{hearingloss1,hearingloss3}.
    To mitigate this issue while maintaining a natural listening experience, open-fit (or semi-open-fit) HAs have been developed \cite{semi-open3,semi-open,open-fit1}. By employing a physical vent to relieve ear canal pressure, these configurations effectively bypass the occlusion effect. However, in such configurations (Figure \ref{fig:leakage path}), \textit{external noisy signal inevitably leaks into the ear canal, leading to degraded BSE performance}.


     \subsection{Related Works}
     \noindent\textbf{Conventional  BSE. }
     Existing BSE approaches can be broadly categorized into two main paradigms: model-driven and data-driven .
     Specifically, model-driven BSE adopts statistical signal processing methods derived from physical models of the degradation and probabilistic models of the involved signals \cite{SPM}, which typically exploit the statistical independence between the target speech and the interference fields. For example, the binaural minimum variance distortionless response (BMVDR) \cite{BMVDR1,BMVDR2}  beamformer (BF) extended the classical MVDR BF \cite{MVDR} to the binaural setting,  which aims to minimize the output noise power while preserving the target source and its spatial cues. Analogously, based on the multi-channel Wiener filter (MWF) \cite{MWF}, several binaural MWF approaches \cite{BMWF1,BMWF2} were presented which extend the mean squared error minimization to the binaural setup, aiming to preserve the spatial perception of the target speech while balancing noise reduction and speech distortion. These model-driven BSE approaches are often computationally efficient and perform well when the ground-truth spatial covariance matrices (SCMs) of speech and noise are available. In practice, however, accurately estimating SCMs is challenging due to the temporal variability of speech and the complexity of real-world acoustic scenes, resulting in substantial performance degradation and even speech distortion in model-driven BSE. With the advent of deep learning, data-driven approaches \cite{wang2025lightweight,gajecki,sun2019deep} have shown tremendous potential for BSE owing to their capability to model complex, non-linear dependencies in acoustic scenes. 
     Among them, mask-based methods \cite{maping1,maping3,maping2,maping4} learned time-frequency masks from noisy inputs to derive essential statistics, e.g., SCMs, or beamforming weights.
     In contrast, end-to-end frameworks \cite{GCRN,duantodaun1,duantodaun2} bypassed the traditional beamforming pipeline by directly learning a mapping from the noisy waveform to the clean binaural signal.
     By exploiting data itself to guide and refine the algorithm design, data-driven methods achieved strong performance in BSE. 
     Nevertheless, they are still constrained by issues such as prohibitive computational complexity for hearing-aid deployment and high sensitivity to train–test mismatch. 
     In addition, most existing BSE approaches, whether model-driven or data-driven, were designed for closed-fit HAs and thus cannot handle the acoustic leakage problem in open-fit HAs (Figure \ref{fig:leakage path}).

    \noindent\textbf{Active BSE (ABSE). }
     To suppress the acoustic leakage in open-fit HAs, active noise control (ANC) \cite{Kuo1999Active,ANC2,ANC3} has been incorporated into the BSE framework, forming a hybrid scheme that is sometimes referred to as ABSE. 
     Its mechanism is that the internal loudspeaker simultaneously plays the enhanced signal and an ``anti-leakage" signal that cancels the leakage component by generating a sound wave with the same amplitude but inverted phase, utilizing the principle of destructive interference. 
     BSE and ANC can be combined in either a cascaded architecture \cite{serizel2010integrated,seriro}  or a parallel architecture  \cite{sabin2024modeling}, with BSE performing speech enhancement and ANC suppressing the leakage signal.
     Alternatively, state-of-the-art methods \cite{xiao2023,10445843} formulated BSE and ANC within a unified optimization problem and derived both closed-form and adaptive solutions, aiming to minimize the global residual noise rather than optimizing each stage individually.
     Nevertheless, these ABSE methods place an error microphone deep inside the ear canal to provide feedback for model-driven adaptive filters, which is difficult in practice because the ear canal is too narrow to accommodate a microphone, making it uncomfortable for long-term wear.
     Although some data-driven ANC approaches \cite{zhang2021deepanc,yaish2025active} can operate without the error microphone, they were not originally designed for HA scenarios and incur high computational costs.
     These limitations emphasize the necessity of developing lightweight and data-driven ABSE methods that do not rely on an error microphone placed deep inside the ear canal.

    \subsection{Contribution}
     In this work, we propose the\textbf{ ABSE-NET}, a lightweight neural model for ABSE in open-fit HAs. 
     The pipeline of ABSE-NET cascades a BMVDR BF  with a lightweight neural network (LNN), thereby combining the benefits of both model-driven and data-driven approaches.
     Specifically, the BMVDR BF performs a coarse BSE  while preserving the spatial cues, whereas the LNN is designed to simultaneously cancel the acoustic leakage and compensate for BMVDR-induced distortion.
     The LNN adopts an encoder–decoder architecture with an intermediate feature augmentation module, in which a novel frequency–time dependency learning (F-TDL) block and a convolutional attention (ConvAtt) block are introduced to effectively refine latent representations.
     To the best of our knowledge, ABSE-NET is the first lightweight ABSE framework without requiring an in-ear error microphone.
     Extensive experiments demonstrate that ABSE-NET outperforms state-of-the-art approaches while exhibiting remarkable superiority in terms of computational efficiency.
     \textit{Notations:} Scalars, vectors, and matrices are represented by lowercase letters, bold lowercase letters, and bold uppercase letters, respectively.

\section{Preliminaries}
    We consider an open-fit HA configuration where each HA consists of $M/2$ microphones, with $M$ being an even number.
    In the short-time Fourier transform (STFT) domain, letting $l$ and $k$ be the time and frequency indices, the multi-channel microphone signal $\boldsymbol{y}(k,l)\in\mathbb{C}^{M\times 1}$ can be represented as
    \begin{align}
        \boldsymbol{y}(k,l) = \boldsymbol{a}(k)s(k,l) + \sum_{i=1}^I\boldsymbol{b}_i(k)n_i(k,l) + \boldsymbol{v}(k,l),
    \end{align}   
    where $\boldsymbol{a}(k)$ and $\boldsymbol{b}_i(k)$ denote the acoustic transfer functions (ATFs) of the target signal $s(k,l)$ and the $i$th interferer $n_i(k,l)$, respectively, $\boldsymbol{v}(k,l)$ denotes the received background noise and sensor self-noise, and $I$ is the number of interferers. For notational simplicity, indices $k$ and $l$ will be omitted from the rest of this paper.
    \begin{figure*}[t] 
        \vspace{-24pt}
        \centering
        \includegraphics[width=1\textwidth, height=1\textheight, keepaspectratio]{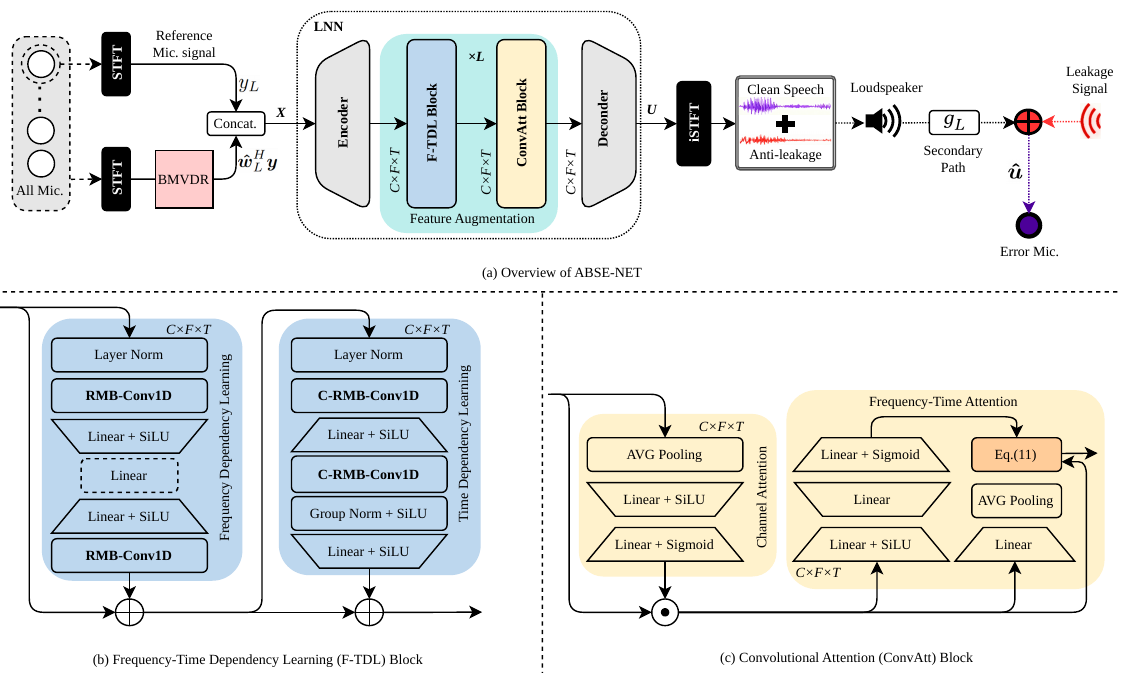}
        \caption{Proposed ABSE-NET: (a) Overview of ABSE-NET, (b) Architecture of F-TDL block, (c) Architecture of ConvAtt block.}
        \vspace{-5pt} 
        \label{fig:module}
    \end{figure*}

    The traditional BMVDR BF \cite{BMVDR1,BMVDR2} consists of two filters designed to preserve the target signal as received by the  reference microphone in each HA, while minimizing the output interferer-plus-noise power. To be specific, taking the left HA as an example, the filter coefficient $\boldsymbol{w}_L$ can be obtained by solving the convex problem:
    \begin{equation} \label{BMVDR}
    \hat{\boldsymbol{w}}_L = \arg \min_{\boldsymbol{w}_L} \boldsymbol{w}^H_L \boldsymbol{R} \boldsymbol{w}_L \quad \text{s.t.} \quad \boldsymbol{w}_L^H \boldsymbol{a} = a_L,
    \end{equation}
    where $\boldsymbol{\hat w}_L$ is the estimate of $\boldsymbol{w}_L$, $\boldsymbol{R}$ is the SCM of the interferer-plus-noise component, superscript $(\cdot)^H$ denotes the Hermitian transpose, and $a_L$ is the ATF from the target source to the reference microphone at the left HA. 
    Based on the Lagrange multiplier method \cite{Lagrange}, $\boldsymbol{\hat w}_L$ can be obtained in closed form, which enables fast computation of the enhanced signal $\boldsymbol{\hat w}_L^H\boldsymbol{y}$.
    In open-fit HAs, using BMVDR BF introduces two main issues:

    \noindent
    \textbf{(1) Acoustic Leakage.} Assume that an error microphone is placed deep inside the ear canal\footnote{\label{fn:two} The error microphone is primarily employed during the signal modeling and model training phases, thereby substantially alleviating the strict dependency on real-time error feedback during inference and deployment.}, its received signal in the STFT domain can be modeled as
    \begin{align} \label{AL_BMVDR}
        e_L = g_L \cdot \boldsymbol{\hat w}_L^H\boldsymbol{y} + d_L \;,
    \end{align}
    where $d_L$ is the acoustic leakage signal in the left HA, and $g_L$ represents the secondary path, i.e., the ATF from the loudspeaker to the error microphone in the left HA. The interferer-plus-noise component in $d_L$ inevitably degrades the output signal-to-interference-plus-noise ratio (SINR), while the target speech component in $d_L$ interacts with $\boldsymbol{\hat w}_L^H\boldsymbol{y}$ and may introduce unpleasant artifacts such as comb filtering \cite{CombFilter}.

    \noindent
    \textbf{(2) Speech Distortion.} In practice, the ground truth of the ATF $\boldsymbol{a}$ is unavailable to the BMVDR BF. As a result, ATF estimation errors lead to a mismatch in the constraint of (\ref{BMVDR}), which in turn causes distortion of both the target speech and the spatial cues. In addition, the SCM $\boldsymbol{R}$ is subject to estimation errors, which degrades the noise reduction performance.

    To tackle the above issues, this work designs a data-driven post-filter for the BMVDR BF, which is mathematically described by the mapping $\mathcal{F}(\cdot)$:
        \begin{align} \label{MAPP_F}
        \boxed{
        \mathcal{F}(\boldsymbol{\hat w}_L^H\boldsymbol{y}, y_L) \mapsto  -d_L/g_L +  a_L s
        }
    \end{align}
    where $y_L$ denotes the signal received by the reference microphone of the left HA. Clearly, the first term aims to cancel the leakage signal, whereas the second term compensates for the distortion in the BMVDR output $\boldsymbol{\hat w}_L^H\boldsymbol{y}$.
    Here, to generate high-quality anti-leakage ``$- d_L/g_L$'',  the noisy reference signal $y_L$ is also fed into the model as an auxiliary input, so as to prevent excessive noise suppression in the BMVDR BF.

\section{Proposed Method}  
    ABSE-NET operates independently on the left and right HAs. For brevity, only the left-HA pipeline is illustrated in Figure \ref{fig:module}, while the right-HA pipeline follows analogously. 
    First, the BMVDR BF performs coarse BSE in the STFT domain, and its output $\boldsymbol{\hat w}_L^H\boldsymbol{y}$ is concatenated with the noisy reference signal $y_L$ to complement the target estimate with sufficient noise information retained in the raw reference and then fed into the LNN.
    The LNN consists of an encoder for feature extraction, a feature augmentation (FA) module, and a decoder that maps the augmented features back to the STFT domain. 
    Finally, the output of LNN is transformed in the time domain, emitted by the loudspeaker, and propagated through the secondary path $g_L$ to the ear canal where it destructively interferes with the leakage signal.

    \noindent 
    \textbf{Symbol Definition.}
    The input $\boldsymbol{X}\in\mathbb{R}^{4\times F \times T}$ of LNN is formed by concatenating the real and imaginary parts of both $\boldsymbol{\hat w}_L^H\boldsymbol{y}$ and $y_L$, where $F$ and $T$ denote the numbers of frequency bins and time frames, respectively. 
    The encoder then extracts the latent feature $\boldsymbol{H}\in\mathbb{R}^{C\times F \times T}$ from $\boldsymbol{X}$, where $C$ is the feature dimension.
    The FA module produces an augmented feature $\boldsymbol{A}\in\mathbb{R}^{C\times F \times T}$, which is transformed to the STFT-domain signal $\boldsymbol{U}\in \mathbb{R}^{2\times F \times T}$ through the decoder.
    The output $\boldsymbol{U}$ of decoder is first transformed into the time domain and then propagated through the secondary path, yielding final output $\boldsymbol{\hat u}$.

    \noindent
    \textbf{Model Details.}
    The BMVDR BF has been detailed in Section 2.
    For the LNN, the encoder employs a single convolutional layer with RMB-Conv1D kernels without identity branches (see subsection 3.1 for details of RMB-Conv1D), while the decoder adopts a fully connected (FC) layer to project the high-dimensional deep features back to the complex spectral dimension.
    The remaining FA module is repeated $L$ times to progressively augment the feature representation through hierarchical abstraction, each consisting of an F-TDL block and a ConvAtt block, whose technical details are described in the following subsections.

    \noindent 
    \textbf{Loss Function.} We define the loss function as
    \begin{align}
       \mathcal{L} = -\text{SI-SDR}(\boldsymbol{\hat u}, \boldsymbol{u}) - \lambda \cdot {\text{STOI}}(\boldsymbol{\hat u}, \boldsymbol{u}),
    \end{align}
    where $\boldsymbol{u}$ denotes the clean speech serving as the ground-truth target and $\lambda$ is a hyperparameter that controls the relative contribution of signal reconstruction quality and perceptual intelligibility. $\text{SI-SDR}(\cdot)$ stands for scale-invariant signal-to-distortion ratio \cite{leroux2019sdr}, designed to optimize waveform reconstruction quality, by measuring the ratio between the target component and the residual distortion while being invariant to global gain scaling, and $\text{STOI}(\cdot)$ represents short-time objective intelligibility \cite{taal2010short}, utilized to enhance speech intelligibility by maximizing the short-time correlation between the temporal envelopes of clean and processed speech in one-third octave bands.

    \subsection{F-TDL Block}
     Speech signals are produced by the periodic vibration of the vocal folds and are subsequently modulated by the vocal tract \cite{theory}, which results in inherent frequency-time dependencies.
     The frequency dependencies stem from: (a) the concentration of speech energy around the fundamental frequency and its harmonics; and (b) the spectral leakage in the STFT domain due to the windowing operation \cite{speech1,speech2,SPEECHal2}.
     The time  dependencies arise from: (a) the short-time quasi-periodicity in the time domain; and (b) the inter-frame convolution caused by room impulse responses longer than the frame length \cite{time1,time2,time3}.
     By exploiting these frequency-time dependencies, state-of-the-art approaches have achieved significant performance improvements in several speech processing tasks \cite{hao,Spatital}.
     However, most of them rely on  computationally intensive attention mechanisms, such as multi-head self-attention in \cite{Spatital}, which makes them ill-suited for deployment in HA platforms.
     To this end, we design the F-TDL block, which consists of two sub-blocks: a frequency dependency learning (FDL) sub-block and a time dependency learning (TDL) sub-block.

     \noindent\textbf{Frequency Dependency Learning Sub-block. }
      This sub-block aims to learn the frequency-dependent information by processing each time frame independently, i.e., weights are shared across time frames.
      As shown in the left part of Figure \ref{fig:module} (b), the FDL sub-block adopts a residual structure \cite{residual} to mitigate the vanishing gradient problem and stabilize the learning dynamics during deep feature extraction. This sub-block consists of layer normalization (LN) followed by a combination of linear (or linear+SiLU) layers and RMB-Conv1D layers.
      LN is applied per sample over all dimensions, ensuring that the normalization is robust to the signal's internal variance and independent of the batch size. Specifically, three linear layers form a bottleneck structure that compresses the feature dimension from \(C\) to \(C_{1}\) and then expands it back to \(C\), where the first and third linear layers are followed by the SiLU nonlinear activation functions \cite{SiLU},  whose smooth and non-monotonic nature helps prevent the dying neuron problem commonly encountered in standard ReLU. The RMB-Conv1D layer applies reparameterized multi branch one-dimensional convolution (RMB-Conv1D) \cite{ding2021repvgg} along the frequency dimension, which employs multi-scale convolutions during training and fuses them into a single kernel at inference.
      In this work, we define three kernel sizes $\mathcal{K}=\{q_1, q_2, q_3\}$ with $q_1 < q_2 < q_3$. Then, in the training stage, the  output $\boldsymbol{Y}_{train}$ of RMB-Conv1D layer can be formulated as
    \begin{align} \label{RMB-Conv1D}
        \boldsymbol{Y}_{train} = \boldsymbol{\Psi } + \sum_{q \in \mathcal{K}} 
        \mathcal{P}(\boldsymbol{\Psi}, \boldsymbol{r}_q),
    \end{align}
    where $\boldsymbol{\Psi }\in\mathbb{R}^{C \times F} $ is the input feature, $\mathcal{P}(\cdot)$ denotes the depth-wise convolution, and $\boldsymbol{r}_q$ is the convolution kernel with $q$ being the kernel size.
    During the inference stage, the output $\boldsymbol{Y}_{infer}$ can be obtained by fusing the multi-scale convolutions in (\ref{RMB-Conv1D}) into a single convolution, i.e.,
    \begin{align} \label{RMB-Conv1D_inference}
        \boldsymbol{Y}_{infer} = \mathcal{P}(\boldsymbol{\Psi}, \boldsymbol{r})
    \end{align}
    where $\boldsymbol{r}$ denotes the fused convolution kernel, which can be calculated by
       \begin{align} \label{fused_kernel} 
        \boldsymbol{r} = \boldsymbol{r}_{q_3} 
        + \mathcal{Z}(\boldsymbol{r}_{q_2}) 
        + \mathcal{Z}(\boldsymbol{r}_{q_1}) 
        + \boldsymbol{r}_I, 
    \end{align}
    where  $\mathcal{Z}(\cdot)$ denotes zero-padding alignment to the maximum kernel size $q_3$, and $\boldsymbol{r}_{I}$ represents the convolution kernel transformed from the first term in (\ref{RMB-Conv1D}). We can conclude from (\ref{RMB-Conv1D_inference}) that only a  single convolution kernel is required during inference stage, while achieving the same modeling capacity as the multi-scale convolutional structure in (\ref{RMB-Conv1D}). In addition, the FDL sub-block avoids computationally expensive multi-head attention mechanisms, enabling efficient deployment on HAs.
   
    \noindent\textbf{Time Dependency Learning Sub-block. }
    This sub-block aims to learn the time-dependent information, i.e., weights are shared across all frequency bins.
    As illustrated in the right part of Figure \ref{fig:module} (b), the TDL sub-block also adopts a residual structure, which is composed of LN followed by a combination of linear+SiLU layers, C-RMB-Conv1D layers, and a group normalization (GN) layer with SiLU activation.
    LN is applied per sample over all dimensions. The C-RMB-Conv1D layer replaces the standard convolution in the RMB-Conv1D layer with causal convolution to enhance the real-time capability of the model by strictly restricting the receptive field to past and present frames, thus preventing any reliance on future look-ahead. Instead of using a bottleneck structure, the feature dimension is first expanded from \(C\) to \(C_{2}\) via a linear+SiLU layer, refined by C-RMB-Conv1D and GN+SiLU, and finally projected back to \(C\) with another linear+SiLU layer.
    This design is motivated by the higher complexity of temporal dependencies in speech signals, which are consequently modeled in a higher-dimensional feature space, thereby preventing the loss of fine-grained temporal dynamics that might occur in a compressed representation. Owing to the use of C-RMB-Conv1D layers, the TDL sub-block maintains a lightweight design while ensuring causality for real-time processing.

\subsection{ConvAtt Block}

    To further refine the features extracted by the F-TDL module, we cascade a ConvAtt block after it. To reduce computational complexity, inspired by~\cite{woo2018cbam}, the ConvAtt sequentially infers attention maps along two separate dimensions, namely the channel dimension and the frequency–time dimension, which avoids the prohibitive computational costs associated with computing full 3D attention tensors, making it highly suitable for lightweight edge deployment.
    Then, the attention maps are applied to the input feature for adaptive feature refinement.

    \noindent\textbf{Channel Attention Sub-block. }
    This sub-block is constructed to produce a channel attention map by exploiting the inter-channel features, which can be formulated as
     \begin{align} \label{Mc}
        \boldsymbol{\Phi}' =  \boldsymbol{M}_C(\boldsymbol{\Phi}) \odot \boldsymbol{\Phi},
    \end{align}   
    where $\boldsymbol{\Phi}\in \mathbb{R}^{C \times F \times T} $ and  $\boldsymbol{\Phi}'\in \mathbb{R}^{C \times F \times T} $ represent the input and output, respectively, $\boldsymbol{M}_C(\boldsymbol{\Phi}) \in \mathbb{R}^{C \times 1 \times 1} $ is the channel attention map, and  $\odot$ denotes the element-wise multiplication with broadcasting.
    As shown in the left part of Figure \ref{fig:module} (c), the channel attention sub-block $\boldsymbol{M}_C(\cdot)$ consists of a global pooling operation over the frequency–time dimensions, followed by a linear+SiLU layer and a linear+Sigmoid layer. 
    The linear+SiLU layer compresses the feature dimension to \(C_3\), while the linear+Sigmoid layer, a linear layer followed by a Sigmoid activation \cite{transformer}, projects the features back to the original dimension \(C\). This design incurs a small number of parameters and therefore maintains a lightweight nature.

    \begin{table*}[t]
        \centering
        \caption{Comparison of computational efficiency, speech quality, and spatial cue preservation on the test set. Note: ``M'' and ``G'' denote millions and gigas, respectively; the best results are highlighted in bold, and the second-best results are underlined.}
        \label{tab:Comparison}
        \small 
        \setlength{\tabcolsep}{6.5pt}
        \renewcommand{\arraystretch}{1.4}
        \begin{tabular}{lrrrrrrrrrr}
            \toprule
            Method & Para.(M) & FLOPs(G) & SI-SDR(dB) & PESQ & STOI & CSIG & CBAK & COVL & $\Delta\text{ILD}$ & $\Delta\text{IPD}$ \\
            \midrule
            Unprocessed   & --             & --                & -2.781            & 1.609             & 0.775             & 2.223             & 1.734             & 1.868             & --                & -- \\
            BMVDR w/o AL  & --             & --                & 5.216             & 3.437             & 0.929             & 4.020             & 3.343             & 3.698             & 4.375             & 0.391 \\
            BMVDR         & --             & --                & 0.878             & 2.196             & 0.861             & 2.670             & 2.266             & 2.424             & 4.054             & 0.351 \\
            FxMWF         & --             & --                & 3.723             & 3.181             & 0.925             & 3.791             & 3.173             & 3.325             & 3.998             & 0.327 \\
            DeepANC       & 19.683         & \underline{5.817} & 2.683             & 2.449             & 0.854             & 3.414             & 2.827             & 2.953             & 3.690             & 0.302 \\
            ASE-TM        & \underline{3.224} & 14.417         & \textbf{10.45}    & \underline{3.573} & \underline{0.953} & \underline{4.212} & \textbf{3.744}    & \underline{4.071} & \underline{3.121} & \textbf{0.242} \\
            ABSE-NET      & \textbf{0.112} & \textbf{0.184}    & \underline{9.869} & \textbf{3.626}    & \textbf{0.955}    & \textbf{4.655}    & \underline{3.578} & \textbf{4.169}    & \textbf{3.047}    & \underline{0.251} \\
            \bottomrule
        \end{tabular}
        \vspace{-5pt}
    \end{table*}


    \begin{figure*}[t] 
        \vspace{0pt}
        \centering
        \includegraphics[width=1\textwidth, height=1\textheight, keepaspectratio]{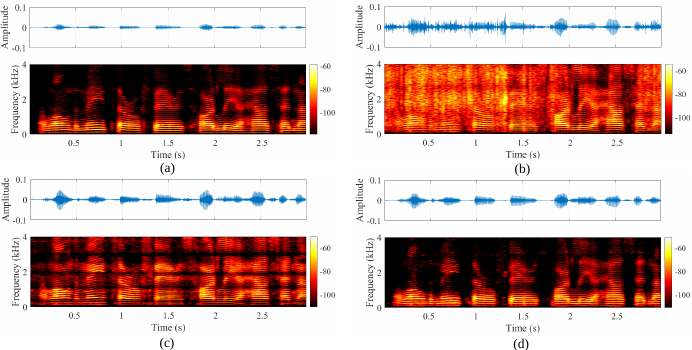}
        \caption{Visualization of waveforms and spectrograms across different signal processing stages: (a) Clean, (b) Unprocessed, (c) BMVDR w/o AL, (d) ABSE-NET.}
        \vspace{-5pt} 
        \label{fig:single_column}
    \end{figure*}

    \noindent\textbf{Frequency-Time Attention Sub-block. } 
    This sub-block produces a frequency-time attention map by exploiting the frequency-time features, which can be formulated as
     \begin{align} \label{Mft}
        \boldsymbol{A} =  \boldsymbol{M}_{FT}(\boldsymbol{\Phi}') \odot \boldsymbol{\Phi}',
    \end{align}   
    where $\boldsymbol{M}_{FT}(\boldsymbol{\Phi}') \in\mathbb{R}^{1 \times F \times T} $ is the frequency-time attention map.
    As shown in the right part of Figure \ref{fig:module} (c), $\boldsymbol{M}_{FT}(\cdot)$ consists of four linear layers (including a linear+SiLU layer and a linear+Sigmoid layer) and an average pooling layer. Mathematically, it can be formulated as
    \begin{align}\label{M_FT}
        \boldsymbol{M}_{FT}(\boldsymbol{\Phi}') = [1 - \alpha \cdot \rho(\boldsymbol{\Phi}') \cdot \boldsymbol{G}(\boldsymbol{\Phi}')] ,
    \end{align}
    where $\boldsymbol{G}(\boldsymbol{\Phi}')\in \mathbb{R}^{1 \times F \times T}$ and $\rho(\boldsymbol{\Phi}') \in \mathbb{R}^{1 \times 1 \times 1}$ are two intermediate variables, and $\alpha$ is a hyperparameter. 
    On the one hand, $\boldsymbol{G}(\boldsymbol{\Phi}')$ is obtained by passing $\boldsymbol{\Phi}'$ through three linear layers. Specifically, the first linear+SiLU layer compresses the channel dimension to $C_4$, the second linear layer projects it back to $C$, and the third linear+Sigmoid layer further reduces the channel dimension to $1$.
    On the other hand, $\rho(\boldsymbol{\Phi}')$ is computed by passing  $\boldsymbol{\Phi}'$ through a linear layer followed by an average pooling layer. The linear layer compresses the channel dimension to 1, while the average pooling layer reduces the frequency-time dimension to 1.
    Similar to the channel attention sub-block, the  frequency-time attention sub-block still maintains the lightweight nature.

\section{Experiments}
  \subsection{Experimental Setups}
   \textbf{Dataset.} The clean speech and noise signals were obtained from the Librispeech dataset \cite{7178964} and the NOISEX-92 dataset \cite{varga1993assessment}, respectively.
   Head-related impulse responses (HRIRs) were derived from the Hearpiece database \cite{Denk2020TheHD}, which provides three external microphones, as well as an in-ear microphone acting as the error microphone. From this database, we selected HRIRs corresponding to 24 distinct incident directions. Specifically, 12 directions were allocated for training, while the remaining 12 were reserved for validation and testing. To ensure adequate spatial coverage and evaluate the model's spatial generalization, the closest adjacent directions within each 12-direction subset were separated by 15°.
   The clean speech and noise signals were convolved with the HRIRs to generate binaural clean speech components and noise components. These two components were then added together at a random signal-to-noise ratio (SNR) ranging from –5 dB to 0 dB to simulate challenging low-SNR conditions typical of daily HA usage, resulting in 43,200 two-second samples (approximately 24 hours in total). The dataset was partitioned into training, validation, and testing sets at an 8:1:1 ratio, with strict mutual exclusivity enforced to ensure absolutely no data overlap across the subsets. All signals were downsampled to 16 kHz.

   \noindent
    \textbf{Training Details.} The STFT was applied using a Hanning window with a length of 320 samples  and a hop size of 160 samples. 
    For the BMVDR beamformer, either ideal or mismatched ATFs were used, while the SCM $\boldsymbol{R}$ was estimated via voice activity detection.
    For the LNN, frame-level normalization was applied to the input $\boldsymbol{X}$ to ensure consistent input scaling, thereby mitigating the impact of drastic energy fluctuations in diverse acoustic environments and stabilizing the learning process. 
    In terms of model architecture, the FA module was repeated $L=4$ times, and all  RMB-Conv1D (or C-RMB-Conv1D) layers included kernel sizes of $\mathcal{K}=\{1, 3, 5\}$.
    The feature dimensions in the FA module were set to $C=16$, $C_1=8$, $C_2=24$, $C_3=4$, and $C_4=4$.
    The hyperparameters $\alpha$ and $\lambda$ were set to 0.3 and 10, respectively.
    The Adam optimizer \cite{adam} was used to train the ABSE-NET with an initial learning rate of $3 \times 10^{-3}$. The batch size was fixed to 6, and the model was trained for 60 epochs, at which point the loss was empirically observed to plateau, ensuring sufficient convergence while avoiding overfitting.
    An in-ear microphone, i.e., the error microphone, was employed in the training stage, but it was not required in the inference stage.

    \noindent   
    \textbf{Evaluation Metrics.} We established a multi-dimensional evaluation framework encompassing speech quality, spatial cue preservation, and computational efficiency to comprehensively assess the practical viability of the proposed ABSE-NET.
    Speech quality was evaluated using the SI-SDR,  perceptual evaluation of speech quality (PESQ) \cite{rix2001perceptual}, STOI,  signal distortion prediction (CSIG), background noise intrusiveness prediction (CBAK), and overall speech quality prediction (COVL) \cite{hu2007evaluation}, thereby enabling a holistic analysis ranging from waveform fidelity to human auditory perception.
    Spatial cue preservation was tested using the interaural level difference (ILD) and interaural phase difference (IPD) errors, as defined in \cite{wang2025lightweight}, which quantify the preservation of directional localization cues essential for spatial awareness.
    Specifically, the ILD errors ($\Delta{\text{ILD}}$) and IPD errors  ($\Delta{\text{IPD}}$) are mathematically formulated as:
   \begin{equation}
    \resizebox{0.88\linewidth}{!}{
        $\begin{aligned}
            \Delta{\text{ILD}} &= \frac{20}{TF} \sum_t \sum_f \left( \log_{10} \left( \frac{|\boldsymbol{\hat u}_L|}{|\boldsymbol{\hat u}_R|} \right) - \log_{10} \left( \frac{|\boldsymbol{u}_L|}{|\boldsymbol{u}_R|} \right) \right) , \\
            \Delta{\text{IPD}} &= \frac{1}{TF} \sum_t \sum_f \left( \arctan \left( \frac{|\boldsymbol{\hat u}_L|}{|\boldsymbol{\hat u}_R|} \right) - \arctan \left( \frac{|\boldsymbol{u}_L|}{|\boldsymbol{u}_R|} \right) \right).
        \end{aligned}$
    }
    \end{equation}
    In addition, computational efficiency was assessed in terms of the number of model parameters and floating-point operations (FLOPs).

    \subsection{Comparison Study}
     To comprehensively evaluate the effectiveness of the proposed ABSE-NET in open-fit HAs, we conducted comparisons with the following baseline methods:
      \begin{itemize}
         \item Model-driven baselines: The BMVDR BF was carried out under two cases, i.e., in the absence and presence of acoustic leakage, with the former referred to as BMVDR w/o AL. In addition, the FxMWF \cite{serizel2010integrated}  was included as another baseline, which enables simultaneous speech enhancement and acoustic leakage suppression using adaptive filtering.
         \item Data-driven baselines: To the best of our knowledge, developing data-driven approaches for the proposed ABSE remains an open problem. Therefore, we extended two closely related data-driven methods to the open-fit HA scenario.
         The first method is DeepANC \cite{zhang2021deepanc}, a deep-learning ANC framework that can preserve target speech by appropriately designing the loss functions.
         The second method is ASE-TM \cite{yaish2025active}, which extends beyond traditional ANC by actively shaping the clean speech. For a fair comparison, both DeepANC and ASE-TM are retrained on our dataset to perform ABSE.
      \end{itemize} 
    From Table~\ref{tab:Comparison}, it can be observed that the speech quality of the unprocessed signals is limited. After processing with the BMVDR BF without acoustic leakage (BMVDR w/o AL), representing an ideal closed-fit scenario where no leakage noise bypasses the HAs, the speech quality improves substantially. For instance, the SI-SDR increases from -2.781 dB to 5.216 dB, while the PESQ score improves from 1.609 to 3.437. 
    In open-fit HAs, however, acoustic leakage is inevitable, resulting in a significant performance degradation of the BMVDR BF, as evidenced by its SI-SDR plummeting from 5.216 dB to a mere 0.878 dB, and PESQ dropping from 3.437 to 2.196. This highlights that methods developed under closed-fit configurations may not generalize well to open-fit HAs.
    By comparison, the FxMWF is designed for open-fit HAs and can jointly achieve speech enhancement and active noise control, thereby outperforming the BMVDR BF. However, its performance is still slightly lower than that of BMVDR w/o AL, likely because its linear filtering mechanism struggles to completely eliminate the leakage signal in complex acoustic scenes.
    For data-driven methods, both DeepANC and ASE-TM outperform the BMVDR beamformer in terms of speech quality, with the latter exhibiting particularly strong performance compared with other baselines, achieving an impressive SI-SDR of 10.45 dB, the highest among all compared methods. 
    The proposed ABSE-NET achieves the best or second-best speech quality across all evaluated metrics, delivering the best performance in PESQ, STOI, CSIG,  COVL, as well as the the second-best results in SI-SDR and CBAK.
    In addition, the ABSE-NET requires only 0.112M parameters and 0.184G FLOPs, which are significantly lower than those of DeepANC and ASE-TM, highlighting its suitability for resource-constrained HA devices.
    This advantage mainly stems from the lightweight FA module designed in ABSE-NET, whereas DeepANC and ASE-TM employ parameter-intensive architectures such as Convolutional Long Short-Term Memory \cite{zhang2021deepanc} or Transformer-Mamba \cite{yaish2025active} structures.
    Moreover, in terms of spatial cue preservation, ABSE-NET achieves the smallest $\Delta{\text{ILD}}$ and the second-smallest $\Delta{\text{IPD}}$, with values of 3.047 and 0.251, respectively, indicating an excellent ability to maintain binaural spatial perception without distorting directional cues.

    For a more intuitive visualization of the speech enhancement results, Figure \ref{fig:single_column} shows a set of waveforms and spectrograms. The unprocessed signal exhibits pronounced acoustic leakage components that obscure speech harmonics. Although BMVDR w/o AL reduces part of the interference, noticeable residual leakage and harmonic distortion remain. In contrast, ABSE-NET more effectively suppresses acoustic leakage while preserving continuous and well-defined harmonic structures that closely resemble those of the clean speech. These visual results further confirm the advantage of ABSE-NET in mitigating leakage-induced artifacts without introducing additional spectral distortion.
    
    \subsection{Ablation Studies}
    \subsubsection{Effectiveness of RMB-Conv1D}
       To validate the effectiveness of RMB-Conv1D layers in ABSE-NET, we replaced the multi-scale convolution in (\ref{RMB-Conv1D_inference}) with a standard depth-wise convolution using a kernel size of 5 (Line 1 of Table \ref{tab:conv_ablation}), which serves as a conventional baseline to highlight the architectural benefits of multi-branch designs.
       Additionally, we adopted multi-scale convolutions with identical kernel sizes, including $\mathcal{K}=\{1, 1, 1\}$, $\mathcal{K}=\{3, 3, 3\}$, and $\mathcal{K}=\{5, 5, 5\}$, to demonstrate that simply increasing the number of parallel branches with identical receptive fields limits the diversity of extracted features.  
       As shown in Table \ref{tab:conv_ablation}, RMB-Conv1D with heterogeneous kernel sizes achieves the best overall performance.
       This is because convolution with different kernel sizes can effectively capture both coarse-grained and fine-grained features, with smaller kernels extracting high-resolution local spectral variations and larger kernels modeling broader contextual dependencies within the acoustic signal.
       Note that all configurations have comparable parameter counts and FLOPs, as the multi-scale convolutions are fused into a single convolution at inference in (\ref{RMB-Conv1D_inference}).

       \begin{table}[t]
            \centering
            \caption{Impact of different convolution strategies.}
            \label{tab:conv_ablation}
            \small 
            \setlength{\tabcolsep}{1.5pt} 
            \renewcommand{\arraystretch}{1.4}
            \begin{tabular}{lrrrrr}
                \toprule 
                Convolution & Para.(M) & FLOPs(G) & SI-SDR(dB) & PESQ & STOI \\
                \midrule 
                
                $\mathcal{K}=\{5\}$       & \underline{0.112} & 0.184             & 7.918             & 3.283             & 0.932 \\
                $\mathcal{K}=\{1, 1, 1\}$ & \textbf{0.111}    & \textbf{0.175}    & 8.134             & 3.475             & 0.945 \\
                $\mathcal{K}=\{3, 3, 3\}$ & \textbf{0.111}    & \underline{0.179} & \underline{8.984} & \underline{3.557} & \underline{0.947} \\
                $\mathcal{K}=\{5, 5, 5\}$ & \underline{0.112} & 0.184             & 8.222             & 3.548             & 0.943 \\
                $\mathcal{K}=\{1, 3, 5\}$ & \underline{0.112} & 0.184             & \textbf{9.869}    & \textbf{3.626}    & \textbf{0.955} \\
    
                \bottomrule 
            \end{tabular}
            \vspace{-5pt}
        \end{table}

    \subsubsection{Effectiveness of FA Module}
    In this part, we tested the effectiveness of FA module through a series of ablation experiments.
    
       \noindent
        \textbf{Effectiveness of F-TDL Block.} In order to assess the validity of the proposed F-TDL block, we first replaced it with a state-of-the-art frequency-time learning model, namely SpatialNet \cite{Spatital}, which was originally designed for speech separation, denoising and dereverberation.
        By comparing the first two rows in Table \ref{tab:ablation_F-TDL}, it can be observed that the parameter counts of F-TDL block and SpatialNet are comparable, but the FLOPs of the F-TDL block are 7 times lower than those of SpatialNet, which is highly advantageous for power-constrained wearable hearing devices. In addition, using F-TDL block delivers superior  speech enhancement performance across all evaluated metrics compared with SpatialNet, yielding an improvement of 1.06 dB in SI-SDR, alongside noticeable gains in both PESQ and STOI.
        Next, we separately removed the FDL and TDL sub-blocks from the F-TDL block, denoted as w/o FDL and w/o TDL, respectively.
        As can be observed, both variants exhibit a significant performance degradation compared with the complete F-TDL block, confirming the necessity of both sub-blocks for speech enhancement.
        In particular, removing the FDL sub-block leads to a more pronounced degradation: compared with F-TDL, the SI-SDR, PESQ, and STOI drop from 9.869 dB, 3.626, and 0.955 to 5.610 dB, 2.527, and 0.819, respectively, suggesting that accurately capturing frequency-dependent spectral features forms the critical foundation for subsequent temporal refinement.
        Finally, we removed the entire F-TDL block and retained only the ConvAtt block for feature augmentation, denoted as w/o F-TDL. As expected, this variant suffers from a severe degradation in speech quality and performs even worse than FxMWF reported in Table \ref{tab:Comparison}, indicating that the attention mechanism alone, without the dedicated spectro-temporal feature extraction provided by F-TDL, is insufficient to handle complex acoustic environments. These comparisons collectively demonstrate that the proposed F-TDL block, as well as each of its sub-blocks, plays a crucial role in the ABSE-NET.

        \begin{table}[t]
            \centering
            \caption{Ablation study of the F-TDL block.}
            \label{tab:ablation_F-TDL}
            \small 
            \setlength{\tabcolsep}{2pt} 
            \renewcommand{\arraystretch}{1.4}
            \begin{tabular}{lrrrrr}
                \toprule
                Architecture & Para.(M) & FLOPs(G) & SI-SDR(dB) & PESQ & STOI \\
                \midrule
                SpatialNet   & 0.119          & 1.313             & \underline{8.809} & \underline{3.597} & \underline{0.951} \\
                F-TDL        & 0.112          & 0.184             & \textbf{9.869}    & \textbf{3.626}    & \textbf{0.955} \\
                w/o FDL      & \underline{0.005} & 0.148          & 5.610             & 2.527             & 0.819 \\
                w/o TDL      & 0.106          & \underline{0.061} & 6.472             & 3.133             & 0.893 \\
                w/o F-TDL    & \textbf{0.001} & \textbf{0.024}    & 1.769             & 2.169             & 0.775 \\
                \bottomrule
            \end{tabular}
            \vspace{-5pt}
        \end{table}

        \begin{table}[t]
            \centering
            \caption{Ablation study of the ConvAtt block.}
            \label{tab:ablation_attention}
            \small
            \setlength{\tabcolsep}{0.9pt}   
            \renewcommand{\arraystretch}{1.4}
            \begin{tabular}{lrrrrr}
                \toprule
                Architecture & Para.(M) & FLOPs(G) & SI-SDR(dB) & PESQ & STOI \\
                \midrule
                CBAM                                      & \textbf{0.110} & 0.193          & 8.128             & 3.123             & 0.935 \\
                ConvAtt                                   & 0.112          & 0.184          & \textbf{9.869}    & \textbf{3.626}    & \textbf{0.955} \\
                w/o CA                                    & \underline{0.111} & 0.183       & \underline{8.273} & \underline{3.419} & \underline{0.941} \\
                w/o FTA                                   & \underline{0.111} & \underline{0.166} & 7.607         & 3.137             & 0.931 \\
                FTA with $\boldsymbol{G}(\boldsymbol{\Phi}')$ & 0.112          & 0.184          & 7.936             & 3.159             & 0.932 \\
                w/o ConvAtt                               & \textbf{0.110} & \textbf{0.164} & 7.441             & 2.939             & 0.928 \\
                \bottomrule
            \end{tabular}
            \vspace{-5pt}
        \end{table}

        \noindent
        \textbf{Effectiveness of ConvAtt Block.} To further evaluate the effectiveness of the ConvAtt block, we first replaced it with a representative convolutional block attention module, namely CBAM  \cite{woo2018cbam}, which infers attention maps along two separate channel and spatial dimensions to reduce computational complexity.
        By comparing the first two rows in Table~\ref{tab:ablation_attention}, it can be observed that the proposed ConvAtt block outperforms CBAM in terms of speech enhancement performance. Meanwhile, although ConvAtt has a slightly higher parameter count than CBAM, it requires fewer FLOPs.
        Next, we separately removed the channel attention sub-block and frequency-time attention sub-block from the ConvAtt block, denoted as w/o CA and w/o FTA, respectively.
        From Table~\ref{tab:ablation_attention}, we can see that both variants perform worse than the complete ConvAtt block, demonstrating the effectiveness of both the channel attention and frequency–time attention sub-blocks.
        Afterward, we substituted the right-hand side of (\ref{M_FT}) with $\boldsymbol{G}(\boldsymbol{\Phi}')$ alone, referred to as FTA with $\boldsymbol{G}(\boldsymbol{\Phi}')$. The resulting performance degradation demonstrates that the complete design in (\ref{M_FT}) is more effective than using $\boldsymbol{G}(\boldsymbol{\Phi}')$ alone, indicating that the complete mechanism better captures complex multi-dimensional dependencies that a single projection might lose.
        Finally, we removed the entire ConvAtt block. Although this slightly reduces the number of parameters and computational cost, it leads to a significant degradation in speech enhancement performance, dropping to 7.441 dB for SI-SDR, 2.939 for PESQ, and 0.928 for STOI, which justifies the minimal computational overhead of 0.02G FLOPs introduced by the attention mechanism. These results confirm the importance of the ConvAtt block in ABSE-NET.

    \noindent
    \textbf{Impact of FA Depth.}
    In Table \ref{tab:ablation_depth}, we evaluated the impact of the FA depth $L$, i.e., the number of stacked F-TDL and ConvAtt blocks in the FA module, to determine the optimal balance between representation depth and computational overhead.
    As expected, both the parameter count and FLOPs increase gradually as $L$ increases, while the speech enhancement performance improves correspondingly.
    For instance, as $L$ increases from 2 to 5, the parameter count increases from  0.108 M to  0.113 M, and the FLOPs increase from 0.094 G to 0.230 G, highlighting a nearly linear scaling in computational cost, whereas the parameter growth is extremely marginal due to the lightweight design of the blocks.
    Meanwhile, SI-SDR, PESQ, and STOI rise to 10.304 dB, 3.708, and 0.959 from  9.327 dB, 3.427, and 0.934, respectively.
    This is because the learning capability of ABSE-NET improves as the number of F-TDL and ConvAtt blocks increases, allowing the network to perform deeper hierarchical feature abstractions and capture more intricate spectro-temporal relationships.
    For deployment on resource-constrained HA platforms, the choice of $L$ should be guided by practical constraints such as processing latency, power consumption, and the available computational capacity, which is why $L=4$ was adopted as the default configuration in our preceding evaluations, providing a highly effective trade-off between performance and efficiency.

        \begin{table}[t]
            \centering
            \caption{Impact of the depth $L$ of FA module.}
            \label{tab:ablation_depth}
            \small 
            \setlength{\tabcolsep}{3pt}
            \renewcommand{\arraystretch}{1.4}
        
            \begin{tabular}{lrrrrr}
                \toprule
                FA Depth & Para.(M) & FLOPs(G) & SI-SDR(dB) & PESQ & STOI \\
                \midrule
                $L=2$ & \textbf{0.108}    & \textbf{0.094}    & 9.327             & 3.427             & 0.934 \\
                $L=3$ & \underline{0.109} & \underline{0.139} & 9.561             & 3.457             & 0.944 \\
                $L=4$ & 0.112             & 0.184             & \underline{9.869} & \underline{3.626} & \underline{0.955} \\
                $L=5$ & 0.113             & 0.230             & \textbf{10.304}   & \textbf{3.708}    & \textbf{0.959} \\
                \bottomrule
            \end{tabular}      
            \vspace{-5pt}
        \end{table}

        \begin{table}[t]
            \centering
            \caption{Impact of ATF errors on BMVDR BF and ABSE-NET.}
            \label{tab:deviation_comparison}
            \small
            \setlength{\tabcolsep}{1pt}  
            \renewcommand{\arraystretch}{1.4}
            \begin{tabular}{lrrrrr}
                \toprule
               DOA Error & SI-SDR(dB) & PESQ & STOI & $\Delta$ILD & $\Delta$IPD \\
                \midrule
                
                BMVDR w/o AL ($0^\circ$) & \underline{5.216} & \underline{3.437} & \underline{0.929} & 4.375             & 0.391 \\
                BMVDR ($0^\circ$)        & 0.878             & 2.196             & 0.861             & \underline{4.054} & \underline{0.351} \\
                ABSE-NET ($0^\circ$)     & \textbf{9.869}    & \textbf{3.626}    & \textbf{0.955}    & \textbf{3.047}    & \textbf{0.251} \\
                \midrule
                
                BMVDR w/o AL ($7.5^\circ$) & \underline{3.428} & \underline{3.254} & \underline{0.917} & 4.443             & 0.397 \\
                BMVDR ($7.5^\circ$)        & 0.112             & 2.017             & 0.855             & \underline{4.094} & \underline{0.351} \\
                ABSE-NET ($7.5^\circ$)     & \textbf{8.218}    & \textbf{3.455}    & \textbf{0.951}    & \textbf{3.647}    & \textbf{0.299} \\
                \midrule
                
                BMVDR w/o AL ($15^\circ$)  & \underline{1.637} & \textbf{3.077}    & \underline{0.874} & 4.688             & 0.415 \\
                BMVDR ($15^\circ$)         & -1.010            & 1.747             & 0.816             & \underline{4.420} & \underline{0.383} \\
                ABSE-NET ($15^\circ$)      & \textbf{6.434}    & \underline{3.001} & \textbf{0.904}    & \textbf{4.191}    & \textbf{0.344} \\
                
                \bottomrule
            \end{tabular}
            \vspace{-5pt}
        \end{table}

    \noindent
    \textbf{Impact of ATF Errors.}
    As we mentioned in Section 2, the ground truth of ATFs is generally unavailable in practice, resulting in speech distortion and the destruction of spatial cues in BMVDR BF.
    In this part, based on the estimated direction of arrival (DOA), we acquired the ATF by querying a personalized head-related transfer function (HRTF) from the user's HRIR dataset.
    In such cases, the DOA error is inevitable due to the dynamic nature of acoustic environments or limitations in practical estimation algorithms, which in turn introduces ATF errors.
    To this end, we investigated the impact of ATF errors by evaluating objective performance under different levels of DOA errors.
    As shown in Table \ref{tab:deviation_comparison}, all metrics decrease with increasing DOA error.
    Notably, BMVDR BF and  BMVDR w/o AL are highly sensitive to the DOA mismatch; when the DOA error reaches 15°, their SI-SDR scores show only marginal improvement compared with the unprocessed signal in Table \ref{tab:Comparison}, while the standard BMVDR even degrades the signal to an SI-SDR of -1.010 dB.
    In terms of speech quality, BMVDR w/o AL outperforms BMVDR BF, as it ignores the acoustic leakage in the ear canal.
    In contrast,  BMVDR BF better preserves spatial cues than BMVDR w/o AL as evidenced by its consistently lower $\Delta\text{ILD}$ and $\Delta\text{IPD}$ values across all test conditions, which can be attributed to the acoustic leakage component containing a portion of the target signal that helps maintain the original spatial information.
    By comparison, ABSE-NET consistently delivers significant speech enhancement performance; it achieves an SI-SDR improvement exceeding 6 dB even under a $15^\circ$ DOA mismatch, maintaining a robust SI-SDR of 6.434 dB alongside a PESQ of 3.001. In addition, ABSE-NET also provides the smallest ILD and IPD errors under all DOA-error conditions, demonstrating its superior ability to preserve binaural spatial perception despite inaccurate ATFs.
    These results further demonstrate that the LNN in ABSE-NET is capable of simultaneously suppressing acoustic leakage and compensating for the distortion introduced by BMVDR.

    \section{Conclusion}

    In this work, we propose ABSE-NET, a novel neural framework for active speech enhancement in open-fit hearing aids.
    By cascading a BMVDR beamformer with a lightweight neural network, ABSE-NET effectively combines the strengths of both model-driven and data-driven paradigms.
    The BMVDR BF first performs coarse speech enhancement while preserving spatial cues.
    Subsequently, the lightweight neural network simultaneously cancels acoustic leakage and compensates for BMVDR-induced distortion, thereby further improving speech quality.
    Specifically, the network architecture comprises an encoder for feature extraction, a novel feature augmentation module to capture the frequency-time dependencies, and a decoder for reconstructing the time-frequency representation. 
    Compared with model-driven approaches, ABSE-NET delivers superior binaural speech enhancement performance without using an in-ear microphone.
    Simultaneously, ABSE-NET outperforms existing data-driven methods in terms of performance and computational efficiency.
    In future work, we plan to further improve the computational efficiency of ABSE-NET and deploy it on practical open-fit hearing aids.

\section{Generative AI Use Disclosure}
    This disclosure clarifies the use of Generative AI tools in this thesis. The author strictly abides by academic integrity, using tools only as auxiliary means. Specifically, they assist in literature review sorting, language expression optimization, and idea brainstorming, without replacing the author’s independent thinking or original conclusions. All AI-generated content has been verified and corrected. The author bears full responsibility for the thesis’s final content, and its use complies with the university’s academic regulations and ethical norms.

\section{Acknowledgment}
    This work was supported in part by the National Natural Science Foundation
    of China under Grants 62361045 and 62201297; and in part by the Natural Science Foundation
    of Inner Mongolia Autonomous Region under Grants 2025QN06018 and 2026QC0225.
    
\bibliographystyle{IEEEtran}
\bibliography{mybib}

\end{document}